\documentclass{article}
\usepackage[a4paper, total={6in, 8in}]{geometry}
\usepackage[font=small,labelfont=bf]{caption}
\usepackage{authblk}
\usepackage{braket}
\usepackage{indentfirst}
\usepackage{amssymb}
\usepackage{hyperref}
\usepackage{cite}
\usepackage{amsmath}
\usepackage{graphicx}
\usepackage{booktabs}
\usepackage{subcaption}
\usepackage{footnote}
\usepackage{fancyhdr}
\usepackage{xcolor}

\title{On the Fermion Level Crossing in the Electroweak Instanton Background}

\author[]{P.~E.~Mogaddam\thanks{p\_eslambolchi@sbu.ac.ir} }
\author[]{S.~S.~Gousheh\thanks{ss-gousheh@sbu.ac.ir}}

\affil[]{\textit{Department of Physics, Shahid Beheshti University, Tehran, Iran}}

\date{\today}

\begin{document}
	\maketitle
	\begin{abstract}
	We investigate fermion level crossing in the electroweak instanton background, taking into account the Euclidean-time dependence of the fermion energy throughout our analysis, from the field equations to the spectral flow of fermion energy levels. Modifying the standard fermion ansatzes, we show that, irrespective of the model parameters, the duration over which the fermion energy spectrum flows from one continuum to another corresponds to exactly one unit change in the Chern--Simons number of the instanton. We further demonstrate that incorporating this time dependence is essential for establishing a one-to-one correspondence between the number of fermion zero modes and the instanton winding number, providing a numerical confirmation of the index theorem in the context of instanton backgrounds.
	\\
	\\ \noindent\textbf{Keywords}: instanton, level crossing, Chern-Simons number, index theorem  
    \end{abstract}

\section{Introduction}\label{sec1}
The process of fermion level crossing refers to the flow of the fermion energies across zero in a varying background field. This phenomenon is particularly important in the context of fermions interacting with topologically nontrivial backgrounds \cite{tHooft:1976rip,tHooft:1976snw, Jackiw:1976pf, Callan:1976je}. In these frameworks, level crossings are directly connected to nontrivial topological properties of the background and lead to remarkable consequences such as fractional fermion number and induced quantum numbers carried by solitons \cite{JackiwRebbi1976,GoldstoneWilczek1981}. 
 Here, we consider fermions in the electroweak (EW) instanton background, whose gauge field is the $SU(2)$ instanton---the soliton-like solution of the Euclidean Yang-Mills equations that is localized in both space and Euclidean-time, and interpolates between vacua characterized by different values of the Chern-Simons number \cite{d370fac2b94a4edca23e7cf28a022b27}. Along this interpolation, the spectrum of the Dirac operator also  evolves continuously, and fermion energies may cross the zero-energy level. The number and chirality of the resulting zero modes are determined by the Atiyah-Singer index theorem \cite{Atiyah:1963zz} or, equivalently, by operator techniques such as those developed in Ref.~\cite{Brown:1977eb}, which yield a single (right-)left-handed zero mode in an (anti-)instanton background. The zero mode of a massless Euclidean Dirac operator in the presence of an instanton of Yang-Mills fields was first given by 't Hooft \cite{tHooft:1976rip,tHooft:1976snw}. 
 
The fermion level crossing becomes even more significant through its connection with anomalies \cite{tHooft:1976rip,tHooft:1976snw,Adler1969,BellJackiw1969} and nonperturbative processes. In particular, the level crossing picture provides an intuitive description of the fermion number violation: as the background gauge field evolves, the spectral flow of the Dirac operator leads to the net creation or annihilation of fermions. This description plays a key role in understanding baryon and lepton number violation in the EW theory not only through instanton tunnelings, but also through sphaleron transitions \cite{Manton1983,Klinkhamer:1984di}, which contribute to the fermion number violation at high temperatures and energies \cite{Kuzmin:1985mm,Ringwald:1989ee,PhysRevD.36.581}. Moreover, Beyond high-energy physics, similar mechanisms appear in condensed matter systems such as polyacetylene, topological insulators, and superconductors, where domain walls or vortices support fermionic zero modes \cite{JackiwSchrieffer1981,SuSchriefferHeeger1979,Kitaev2001}. In these systems, spectral flow and level crossing underlie phenomena such as charge fractionalization and Majorana bound states, demonstrating the broad applicability of these ideas across different areas of physics.
 
However, it is well known that a gauge theory with scalars has no non-trivial solutions in four-dimensional Euclidean space-time due to the Derrick theorem \cite{Derrick:1964ww}. As a result, in the EW theory, the standard instanton solutions of the field equations are ruled out, unless the instanton size $\rho$ is much smaller than the inverse Higgs vacuum expectation value, $v^{-1}$. In this regime, a constrained instanton \cite{Affleck:1980mp} may still exist and give rise to a fermion zero mode. The process of fermion level crossing in the background of the EW constrained instanton was studied by Ref.~\cite{Yang:1993qi} (and of the EW sphaleron by Ref.~\cite{Kunz_1994}), observing that the bound state exists only in a finite $x_0$-interval, the duration of which depends on the dimensionless variable $\tilde{\rho}=\rho m_f$, where $m_f$ refers to the fermion mass generated after symmetry breaking. Consequently, from the plot of the fermion energies with respect to the instanton Chern-Simons number they conclude that the difference between the Chern-Simons number for the two critical, where the fermion bound state merges into the continua, tends to zero for small values of $\tilde{\rho}$ and tends to one for large values of $\tilde{\rho}$ \cite{Yang:1993qi}. That is, the number of fermion zero modes is not, in general, equal to the instanton number, calling into question their one-to-one correspondence implied by the index theorem. However, this conclusion is obtained by solving the eigenvalue equation for the bound state, ignoring the Euclidean-time dependence of the fermion energy. Although neglecting this dependence is a common practice, we show that it cannot be done consistently in this case. In this paper we present a modification which amounts to taking this Euclidean-time dependence into account consistently throughout the calculations. We then show that, with this refinement, a bound state emerges from the positive continuum at $x_0=-\infty$ and merges with the negative continuum at $x_0=\infty$, regardless of the value of parameter $\tilde{\rho}$. Hence, the conclusions change considerably.
   
The rest of the paper is organized as follows. In Sec.~\ref{sec2}, we introduce the model and establish our notation in the Euclidean space-time. Section~\ref{sec3} is devoted to a detailed analysis of the fermion level crossing process in the instanton background, both with ignoring and taking into account the Euclidean-time dependence of the fermion energy. The former confirms the result of Ref.~\cite{Yang:1993qi}, while the latter shows that the conclusion reached are modified. Finally, our conclusions are summarized in Sec.~\ref{sec4}.
\section{The Model} \label{sec2}
The EW theory is built upon the gauge symmetries $SU(2)_{L}$ and $U(1)_{Y}$, with the corresponding gauge fields 
$A_{\mu}^{a}$ $(a=1,2,3)$ and $B_{\mu}$, whose field strength tensors are defined by
\begin{align}
	F_{\mu\nu}^{a}
	&=\partial_{\mu}A_{\nu}^{a}-\partial_{\nu}A_{\mu}^{a}
	-g\,\varepsilon^{abc}A_{\mu}^{b}A_{\nu}^{c}, \label{F} \\
	\mathcal{B}_{\mu\nu}
	&=\partial_{\mu}B_{\nu}-\partial_{\nu}B_{\mu}. \notag
\end{align}
The fermionic sector of this theory consists of the left-handed doublets $L^i=\{E_L^i,Q_L^i\}$ with leptons $ E_{L}^i=(\nu_{e^i},\,e^i)_{L}^{T}$ and quarks $Q_{L}^i=(u^i,\,d^i)_{L}^{T}$, as well as the right-handed singlets $ R^i=\{e_{R}^i,\,u_{R}^i,\,d_{R}^i\}$, for $\mathbf{e}=(e,\mu,\tau),\, \mathbf{u}=(u,c,t)$ and $\mathbf{d}=(d,s,b)$. 

With the above conventions, the EW Lagrangian in Euclidean space can be expressed in the following compact form \cite{Gibbs:1995xt} (details are shown in App.~\ref{app}):
\begin{align}
	\mathcal{L_\mathrm{EW}}=&\frac{1}{4}\mathcal{B}^{\mu \nu} \mathcal{B}_{\mu \nu}+\frac{1}{4}F^{\mu \nu}_a F_{\mu \nu}^{a} +D_{\mu} \phi^{\dagger} D^{\mu}\phi+\lambda\left(\phi^{\dagger} \phi -\frac{v^{2}}{2} \right)^{2}\notag\\
	&+\bar{R}^i\gamma^{\mu}\mathcal{D}_\mu R^i+\bar{L}^i \gamma^{\mu} D_{\mu}L^i  
	-\bar{E}_L^i y_l^{ij}\phi\, e_R^j - \bar{Q}_{L}^iy_q^{ij}\left(\tilde{\phi}u_{R}^j+\phi d_{R}^j\right)+\mathrm{H.C.} , \label{L}  
\end{align}
where $\tilde{\phi}=i\tau^{2}\phi^{*}$, and the covariant derivative acting on right-handed fermions is given by
\begin{align*}
	\mathcal{D}_{\mu}=\partial_{\mu}-i g' Y B_{\mu},
\end{align*}
while for left-handed fermions and for the Higgs doublet $\phi$ it takes the form
\begin{align*}
	D_{\mu}=\partial_{\mu}- ig A_{\mu}^{a}\frac{\tau^{a}}{2}-i g' Y B_{\mu},
\end{align*}
in which $\{\tau^{a}\}$ denote the isospin Pauli matrices in the fundamental representation, while in the adjoint representation they are given by $(\tau^{b})^{ac} =2i\varepsilon^{abc}$.

In this study, we consider the limit of the vanishing mixing angle, and restrict our analysis to the up and down quarks, which are assumed to be degenerate in mass after symmetry breaking, allowing us to introduce a common mass parameter $m_q = y_q v / \sqrt{2}$.
 We also use the approximation that fermion back-reactions on the gauge and Higgs sectors are negligible, an assumption well justified for classical solitons (see, for example, \cite{Shahkarami_2011}). With these simplifications, the field equations derived from the Lagrangian
in Eq.~\eqref{L} can be written as
\begin{subequations}
	\begin{align}
		&D_{\nu}^{ac} F^{\mu\nu}_c =i\frac{g}{2} \left[\phi^\dagger\tau^a D^\mu\phi-(D^\mu\phi)^\dagger\tau^a\phi\right],\label{aeom}\\&
		D^\mu(D_{\mu}\phi^{a}) =2 \lambda \left(\phi^{\dagger} \phi-\frac{v^2}{2}\right)\phi^{a},\label{phieom}   
		\\& D_{0} Q_{L}+i \sigma^{i} D_{i} Q_{L}=y_q\left(\tilde{\phi} u_{R}+\phi d_{R}\right), \label{qLeom}\\
		& \partial_{0} u_{R} - i \sigma^{i} \partial_{i} u_{R}=y_{q} \tilde{\phi}^{\dagger} Q_{L}, \label{uReom} \\
		& \partial_{0} d_{R}-i \sigma^{i} \partial_{i} d_{R}=y_q \phi^{\dagger} Q_{L}. \label{dReom}
	\end{align} \label{eom}
\end{subequations}
Having ignored the fermion back-reactions, we should first solve Eqs.~\eqref{aeom} and \eqref{phieom},  then substitute the resulting gauge and Higgs field configurations into Eqs.~\eqref{qLeom}--\eqref{dReom}. 

For $v=0$, Eqs.~\eqref{aeom} and \eqref{phieom} admit the trivial solution $\phi(x)=0$, and reduce to the classical Euclidean field equations of the pure SU(2) Yang-Mills theory with the instanton solution (see \cite{rajaraman1982solitons,manton2004topological} for details). An explicit instanton is given in the regular gauge by 
	\begin{align}
	A_\mu(x)=\frac{1}{g}\frac{1}{x^2+\rho^2}\eta_{\mu\nu a}x^\nu\tau^a,\qquad \label{A} 
\end{align}
  where $x^2=x_0^2+\mathbf{x}^2$, $\rho$ is some arbitrary scale parameter often referred to as the instanton size, and $\eta_{\mu\nu a}$
 is the standard 't Hooft symbol \cite{tHooft:1976rip,tHooft:1976snw}. The pure SU(2) Yang-Mills theory has an infinite number of degenerate and topologically distinct
 	vacua, separated by energy barriers and characterized by integer values of the Chern-Simons number,
	\begin{align*}
		N_\mathrm{CS} =\int\mathrm{d^3}x \, K^0(x), 
	\end{align*}
with the topological current
	\begin{align*}
		K^\mu(x)= \frac{g^2}{32\pi^2}\epsilon^{\mu\nu\rho\sigma}\left(F_{\nu\rho}^a(x)A_\sigma^a(x)-\frac{g}{3}\epsilon_{abc}A_\nu^a(x)A_\rho^b(x)A_\sigma^c(x)\right).
	\end{align*}   
The instanton carries a Pontryagin index or winding number $Q = \int \mathrm{d}S_\mu\, K^\mu(x)$ of unity and is commonly interpreted as a field configuration that propagates in imaginary time from a vacuum at $x_0=-\infty$ to its adjacent vacuum at $x_0=\infty$, {\it i.e.,} $Q=N_{\mathrm{CS}}(x_0=\infty)-N_{\mathrm{CS}}(x_0=-\infty)$.  

For $v\ne0$, however, there does not exist a minimum-action solution to the coupled Higgs and gauge field equations of motion, due to the Derrick non-existence theorem \cite{Derrick:1964ww}. Nevertheless, as long as $\rho v\ll1$, the RHSs of Eqs.~\eqref{aeom} and \eqref{phieom} tend to zero\footnote{ They are 
	$-i\frac{g}{2}(\rho v )^2\frac{x^a}{(x^2+\rho ^2)^{2}}$ and $-\lambda(\rho v)^2\frac{x_0+i\mathbf{x}\cdot\boldsymbol{\tau}}{(x^2+\rho^2)^{3/2}}\frac{v}{\sqrt{2}}\binom{0}{1}$, respectively.}, 
and a constrained instanton \cite{Affleck:1980mp} may still exist with the gauge field configuration given by Eq.~\eqref{A} and the Higgs field configuration \cite{tHooft:1976snw}
 \begin{align}
\phi=\frac{x_0+i\mathbf{x}\cdot\boldsymbol{\tau}}{\sqrt{x^2+\rho^2}}\frac{v}{\sqrt{2}}\binom{0}{1}. \label{phi}
\end{align}
In what follows, we investigate the fermion level crossing process in the background field of the EW constrained instanton specified by the configurations in the form of Eqs.~\eqref{A} and \eqref{phi}.

\section{The Level Crossing Process}\label{sec3}
 The process of fermions passing through the zero-energy level in the background field of the EW instanton can be explored by solving the fermion equations of motion, which are given by Eqs.~\eqref{qLeom}--\eqref{dReom}.            
First of all, we use the gauge transformation $\mathbf{V}(x)=\exp\left(i\boldsymbol{\tau}\cdot\mathbf{\hat{x}}\,f(x)\right)$ to set $A_\mu(x)$ into the temporal gauge,
	\begin{align}
		A'_0(x)=\mathbf{V}^{-1}(x)A_0(x)\mathbf{V}(x)+\frac{i}{g}\mathbf{V}^{-1}(x)\partial_0\mathbf{V}(x)=0,\label{temgauge}
	\end{align}
with the corresponding transformation,
	\begin{align}
		\mathbf{A'}(x)=\mathbf{V}^{-1}(x)\mathbf{A}(x)\mathbf{V}(x)+\frac{i}{g}\mathbf{V}^{-1}(x)\boldsymbol{\nabla}\mathbf{V}(x),\label{vecA'}
	\end{align}
under which the scalar field $\phi$ transforms as
    	\begin{align}
		\phi'(x)=\mathbf{V}^{-1}(x)\phi(x).\label{phi'}
	\end{align}
The temporal gauge is preferred because it permits a clear tracking of topological transitions as $x_0$ varies.
     From Eq.~\eqref{temgauge}, the solution for $f$ can be found by integration \cite{Yang:1993qi}
	\begin{align}
		f(x)=\frac{r}{\sqrt{r^2+\rho^2}}\left[\arctan\left(\frac{x_0}{\sqrt{r^2+\rho^2}}\right)+\Theta(r) \right], \label{f}
	\end{align}
	where $r=\sqrt{\mathbf{x}^2}$ and $\Theta(r)$ is the Euclidean-time independent residual gauge freedom which we set to zero. 
	
    Solving Eqs.~\eqref{vecA'} and \eqref{phi'} for the field configurations given by Eqs.~\eqref{A} and \eqref{phi} then yields the following expressions for the gauge and Higgs field \cite{Yang:1993qi}
	\begin{align}
		&\mathbf{A'}(x)=\frac{1}{g}\left[a(x)(\boldsymbol{\tau}\times\mathbf{\hat{x}})+b(x)(\boldsymbol{\tau}-(\boldsymbol{\tau}\cdot\mathbf{\hat{x}})\mathbf{\hat{x}})+c(x)(\boldsymbol{\tau}\cdot\mathbf{\hat{x}})\mathbf{\hat{x}}\right], \notag \\
       &\phi'(x)= \frac{v}{\sqrt{2}}[h(x)+
       i\boldsymbol{\tau}\cdot\mathbf{\hat{x}}\,k(x
       )]\binom{0}{1},
       \label{A'}
	\end{align}
	with the profile functions
	\begin{subequations}
	\begin{align}
			a(x)&=-\frac{1}{x^2+\rho^2}(r \cos{2f}+x_0 \sin{2f})-\frac{\sin^2{f}}{r},  \\
			b(x)&=-\frac{1}{x^2+\rho^2}(x_0\cos{2f}-r\sin{2f})-\frac{\sin{2f}}{2r},  \\
			c(x)&=-\frac{x_0}{x^2+\rho^2}-\partial_r f, \\
			h(x)&=\frac{1}{\sqrt{x^2+\rho^2}}(-r \sin{f}+x_0 \cos{f}),  \\
			k(x)&=\frac{-1}{\sqrt{x^2+\rho^2}}(r \cos{f}+x_0 \sin{f}).
	\end{align} \label{abc}  
	\end{subequations}

\begin{figure}[ht]
 	\begin{subfigure}{.49\textwidth}
 		\centering
 		\includegraphics[width=1\linewidth]{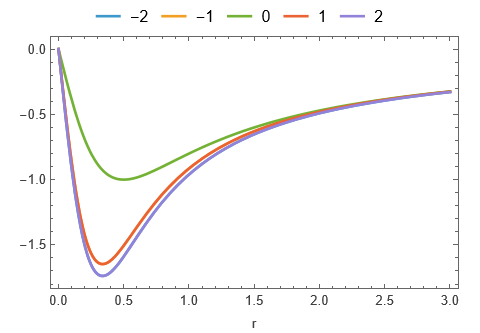}
 		\caption{} \label{apro}
 	\end{subfigure} \hfill
 	\begin{subfigure}{.49\textwidth}
 		\centering	\includegraphics[width=1\linewidth]{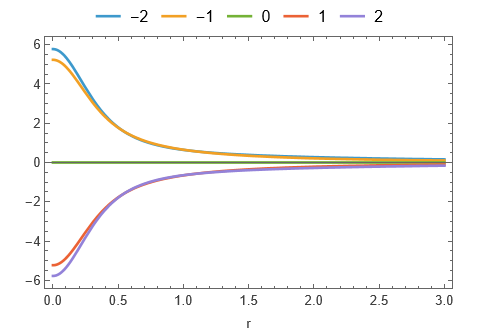}
 		\caption{} \label{bpro}
 	\end{subfigure} \hfill
 	\begin{minipage}[c]{0.49\textwidth}
 \begin{subfigure}{\textwidth}
 \centering
 \includegraphics[width=1\linewidth]{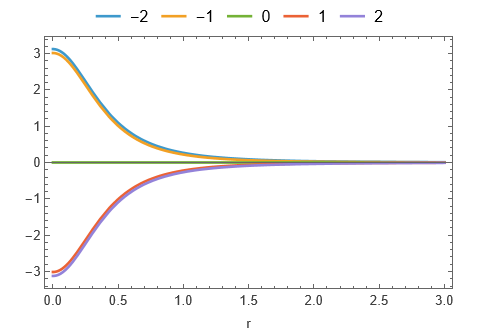}
 \caption{} \label{cpro}
		\end{subfigure}
\end{minipage}\hfill
\begin{minipage}[c]{0.47\textwidth}
 	\caption{The profile functions of the instanton gauge field, (a) $a(x_0,r)$, (b) $b(x_0,r)$ and (c) $c(x_0,r)$, in terms of $r$ for $\rho=0.5$ and $x_0=0, \pm 1,\, \pm2$ in arbitrary units. The first two plots of $a$ overlap with the last two.} \label{abcpro}
 \end{minipage}
 \end{figure}

\begin{figure}[ht]
 	\begin{subfigure}{.49\textwidth}
 		\centering
 		\includegraphics[width=1\linewidth]{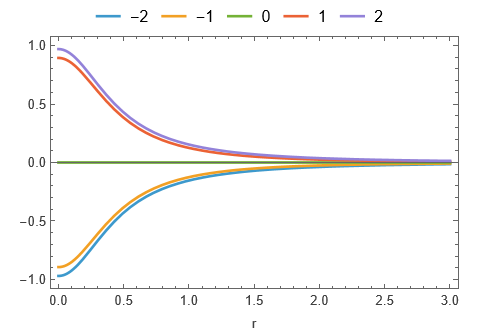}
 		\caption{} \label{hpro}
 	\end{subfigure} \hfill
 	\begin{subfigure}{.49\textwidth}
 		\centering	\includegraphics[width=1\linewidth]{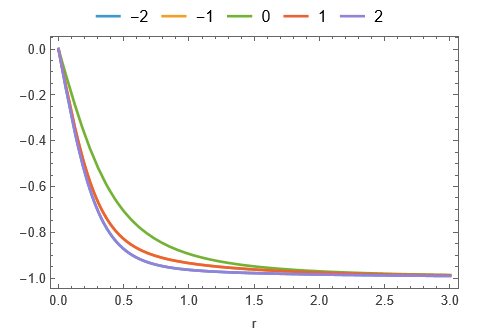}
 		\caption{} \label{kpro}
 	\end{subfigure}
 	\caption{The profile functions of the Higgs field, (a) $h(x_0,r)$ and (b) $k(x_0,r)$, in terms of $r$ for $\rho=0.5$ and $x_0=0, \pm 1,\, \pm2$ in arbitrary units. The first two plots of $k$ overlap with the last two.} \label{hkpro}
 \end{figure}

We illustrate the dynamics of the instanton profile functions $a(x_0,r)$, $b(x_0,r)$, and $c(x_0,r)$ by plotting them in Fig.~\ref{abcpro} as functions of $r$, tracking their evolution from $x_0=-2$ to $x_0=2$ in five steps, for a fixed $\rho$.  The corresponding Higgs field profiles, {\it i.e.}, $h(x_0,r)$ and $k(x_0,r)$, are shown in Fig.~\ref{hkpro}. In view of these figures, the only nonvanishing profile functions at $x_0=0$ are $a$ for the gauge field and $k$ for the Higgs field. The importance of this observation lies in the fact that, as we shall see, the fermions cross the zero-energy level precisely at $x_0=0$. At this point, the four dimensional Euclidean space is effectively restricted to $\mathbb{R}^3$, which can be compactified to a two-sphere $S^2_{\infty}$, while the Higgs field given by Eq.~\eqref{phi} reduces to a Lie-algebra-valued configuration. With three degrees of freedom associated with the isospin Pauli matrices and the constraint $\langle\phi\rangle \to v/\sqrt{2}$ at spatial infinity, the Higgs field now defines a map $\phi : S^2_{\infty} \to S^2$. This feature restricts the constrained instanton to a non-Abelian monopole-like configuration~\cite{tHooft:1974kcl, Polyakov:1974ek}, whose nontrivial second homotopy group ensures the existence of fermion zero modes via the index theorem.

\begin{figure}[ht]	
\centering
\includegraphics[width=0.49\textwidth]{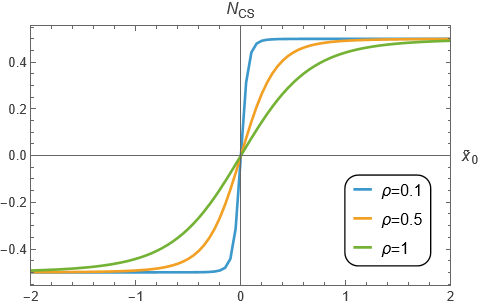}
\caption{ The Chern-Simons number $N_\mathrm{CS}$ as a function of Euclidean time $x_0$ for different instanton sizes $\rho$=0.1 (blue line), 0.5 (orange line) and 1 (green line) in arbitrary units. }
\label{Ncsx0}
\end{figure}

Utilizing Eq.~\eqref{abc}, the Chern-Simons number associated with Eq.~\eqref{A'} becomes \cite{Yang:1993qi}
	\begin{align}
		N_\mathrm{CS}(x_0)=\frac{2}{\pi}\int_0^\infty r^2\mathrm{d}r\left[2c\left(a^2+b^2+\frac{a}{r}\right)+ba'-ab'\right],
  \label{ncs}
	\end{align}
 which is displayed for three different values of the instanton size in Fig. \ref{Ncsx0}.
As shown in this figure, the slopes of the curves near the origin are inversely related to the instanton size. We should note that choosing $\Theta=\pi/2$, for example, yields a Chern--Simons number ranging from 0 to 1.

 To solve the fermion bound state problems, Eqs.~\eqref{qLeom}--\eqref{dReom}, we consider the following ansatzes \cite{Yang:1993qi}
 \begin{subequations}
        \begin{align}
			&Q_{L}(x)=e^{-i\varepsilon x_0}[G_{L}(r)+i \boldsymbol{\sigma}\cdot\mathbf{\hat{x}}\, F_{L}(r)]\chi_h, \\ 
            &u_{R}(x)=\frac{1}{\sqrt{2}}e^{-i\varepsilon x_0} [G_{R}(r)+i \boldsymbol{\sigma}\cdot\mathbf{\hat{x}}\, F_{R}(r)]\chi_1, \\ 
            &d_{R}(x)=\frac{1}{\sqrt{2}}e^{-i\varepsilon x_0} [G_{R}(r)+i \boldsymbol{\sigma}\cdot\mathbf{\hat{x}}\, F_{R}(r)]\chi_2,
		\end{align} \label{ansatz}
\end{subequations}
	with
	\begin{align*}
		\chi_h=\frac{1}{\sqrt{2}}\left[ \begin{pmatrix} 1\\0 \end{pmatrix}_S\begin{pmatrix} 0\\1 \end{pmatrix}_I-\begin{pmatrix} 0\\1 \end{pmatrix}_S\begin{pmatrix} 1\\0 \end{pmatrix}_I\right],\notag \qquad \chi_1=-\begin{pmatrix} 0\\1 \end{pmatrix}_S, \qquad 	\chi_2=\begin{pmatrix} 1\\0 \end{pmatrix}_S,
	\end{align*}
where $S$ refers to spin, $I$ to isospin, and $\chi_h$ is the hedgehog spinor satisfying the spin-isospin relation 
\begin{align}
	(\boldsymbol{\tau}+\boldsymbol{\sigma})\,\chi_h=0, \label{spinisospin}
\end{align} 
$\{\sigma^ i\}$ denoting the Pauli matrices associated with spin. 

To proceed, we first make the usual assumption that $\varepsilon$ is $x_0$-independent, and subsequently show that it leads to inconsistent results. In particular, the prominent inconsistency is that, at the end of the calculations, $\varepsilon$ turns out to be $x_0$-dependent. One may have hoped $\varepsilon$ would be slowly varying with $x_0$, but this will not be the case either. We then generalize the usual procedure by allowing it to be $x_0$-dependent from the start and show that the altered results are self-consistent and the conclusions change significantly.

Using Eqs.~\eqref{A'} and \eqref{ansatz} in Eqs.~\eqref{qLeom}--\eqref{dReom}, four coupled first-order differential equations are obtained \cite{Yang:1993qi} 
\begin{subequations}
\begin{align}
&\tilde{G}_{R}'+(\tilde{h}\tilde{F}_{L}+\tilde{k} \tilde{G}_{L})= \tilde{\varepsilon}\, \tilde{F}_{R}, \\
&\tilde{F}_{R}'+\frac{2}{\tilde{r}} \tilde{F}_{R}-(\tilde{h}\tilde{G}_{L}-\tilde{k}\tilde{F}_{L})= -\tilde{\varepsilon}\, \tilde{G}_{R}, \\
&\tilde{G}_{L}'-2\tilde{a}\tilde{G}_{L}+(2\tilde{b}-\tilde{c})\tilde{F}_{L}-(\tilde{h}\tilde{F}_{R}-\tilde{k}\tilde{G}_{R})=- \tilde{\varepsilon}\, \tilde{F}_{L}, \\
&\tilde{F}_{L}'+(2\tilde{a}+\frac{2}{\tilde{r}})\tilde{F}_{L}+(2\tilde{b}+\tilde{c})\tilde{G}_{L}+(\tilde{k}\tilde{F}_{R}+\tilde{h}\tilde{G}_{R})= \tilde{\varepsilon}\, \tilde{G}_{L}.
\end{align} \label{indepeignequ}  
\end{subequations}
Here, $\tilde{\varepsilon}=\varepsilon/m_q$, $\tilde{r}=m_q r$, $\tilde{G}(\tilde{r})=G(r)/{m_q}^{3/2}$, $\tilde{F}(\tilde{r})=F(r)/{m_q}^{3/2}$, $\tilde{a}(\tilde{r})=a(r)/m_q$, $\tilde{b}(\tilde{r})=b(r)/m_q$, $\tilde{c}(\tilde{r})=c(r)/m_q$, $\tilde{k}(\tilde{r})=k(r)$ and $\tilde{h}(\tilde{r})=h(r)$, where $\tilde{x}_0=m_q x_0$ and $\tilde{\rho}=m_q\rho$.
There are two independent parameters $\tilde{x}_0$ and $\tilde{\rho}$ in these eigenvalue equations. 
Given a value of the parameter $\tilde{\rho}$, we numerically solve Eq.~\eqref{indepeignequ} for different values of the parameter $\tilde{x}_0$, subject to the appropriate boundary conditions. 
 
To extract these conditions, we analyze the  behavior of the equations in the limit $\tilde r \to 0$, while imposing that all wave functions
 $\tilde{G}_{L}$, $\tilde{F}_{L}$, $\tilde{G}_{R}$, and $\tilde{F}_{R}$
 vanish as $\tilde r \to \infty$ in order to have normalizable bound states. Near $\tilde{r}=0$, Eq.~\eqref{indepeignequ} reduces to
 \begin{subequations}
 	\begin{align}
 		&\tilde{G}_{R}'+\frac{\tilde{x}_0}{\sqrt{\tilde{x}_0^2+\tilde{\rho}^2}}\tilde{F}_{L}= \tilde{\varepsilon}\, \tilde{F}_{R}, \label{GR'} \\
 		&\tilde{F}_{R}'+\frac{2}{\tilde{r}} \tilde{F}_{R}-\frac{\tilde{x}_0}{\sqrt{\tilde{x}_0^2+\tilde{\rho}^2}}\tilde{G}_{L}= -\tilde{\varepsilon}\, \tilde{G}_{R}, \label{FR'}\\
 		&\tilde{G}_{L}'-\frac{\tilde{x}_0}{\tilde{x}_0^2+\tilde{\rho}^2}\tilde{F}_{L}-\frac{\tilde{x}_0}{\sqrt{\tilde{x}_0^2+\tilde{\rho}^2}}\tilde{F}_{R}=- \tilde{\varepsilon}\, \tilde{F}_{L}, \label{GL'} \\
 		&\tilde{F}_{L}'+\frac{2}{\tilde{r}}\tilde{F}_{L}-\frac{3\tilde{x}_0}{\tilde{x}_0^2+\tilde{\rho}^2} \tilde{G}_{L}+\frac{\tilde{x}_0}{\sqrt{\tilde{x}_0^2+\tilde{\rho}^2}}\tilde{G}_{R}= \tilde{\varepsilon}\, \tilde{G}_{L}. \label{FL'}
 	\end{align}
 	\label{originlim}
 \end{subequations}
 
 For $\tilde{x}_0=0$, the system of equations~\eqref{originlim} can be decoupled and expressed as a set of second-order differential equations:
 \begin{align}
 	\tilde{G}''_{L/R}
 	+\frac{2}{\tilde{r}}\tilde{G}_{L/R}'=-\tilde{\varepsilon}^2\,\tilde{G}_{L/R}, \qquad \tilde{F}''_{L/R}
 	+\frac{2}{\tilde{r}} \tilde{F}_{L/R}'-\frac{2}{\tilde{r}^2}\tilde{F}_{L/R}=-\tilde{\varepsilon}^2\,\tilde{F}_{L/R}.\label{originGL2}
 \end{align}
   Considering the leading-order polynomial approximation of the wave functions,
   \begin{align}
   	\tilde{G}_{L/R}=\pm c_1\, \tilde{r}^n,\qquad \tilde{F}_{L/R}=\mp c_2\, \tilde{r}^l,\label{poly}
   \end{align}
where $c_1\geq 0$ and $c_2\geq 0$, Eq.~\eqref{originGL2} leads to
   $n(n+1)=0$ and $l(l+1)-2=0$, yielding $n=0,-1$ and $l=1,-2$. Requiring regularity of the fermion wave functions permits only the asymptotic behaviors
   \[\tilde{G}_{L/R}=\pm c_1, \qquad \tilde{F}_{L/R}=\mp c_2 \tilde{r}. \]
 Inserting the above solutions into the original first-order system, Eqs.~\eqref{GR'} and \eqref{GL'} yield
 $3c_2 = -\tilde{\varepsilon} c_1$, while
 Eqs.~\eqref{FR'} and \eqref{FL'} give $\tilde{\varepsilon} c_2 \tilde r = 0$, which implies $ \tilde{\varepsilon}c_2= 0$. Thus, the nontrivial solution with $c_1 \ne 0$, must have $c_2=0$ and $\tilde{\varepsilon} = 0$. This condition is in fact expected, as it removes the ambiguity in defining $\hat{\mathbf{x}}$ at the radial origin that arises from the spinor structure $\boldsymbol{\sigma}\!\cdot\!\hat{\mathbf{x}}$ in the fermion ansatz. Moreover, it is fully supported by the fact that the grand-spin of fermion zero modes in topologically nontrivial backgrounds is zero \cite{JackiwRebbi1976,PhysRevD.26.2058, PhysRevD.45.2920}: since the $\tilde{F}_{L/R}$ components are responsible for generating orbital angular momentum, their absence restricts the system to $\mathbf{L}=0$, extending  Eq.~\eqref{spinisospin} to the familiar grand-spin relation
 $\mathbf{T}+\mathbf{S}+\mathbf{L}=0$, where $\mathbf{T}=\boldsymbol{\tau}/2$ and $\mathbf{S}=\boldsymbol{\sigma}/2$.
 
\begin{figure}[ht]
 	\begin{subfigure}{.48\textwidth}
 		\centering
 		\includegraphics[width=1\linewidth]{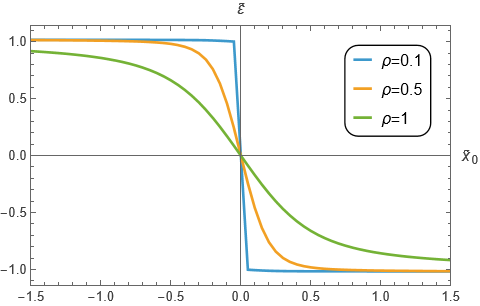}
 		\caption{} \label{ex0}
 	\end{subfigure} \hfill
 	\begin{subfigure}{.49\textwidth}
 		\centering	\includegraphics[width=1\linewidth]{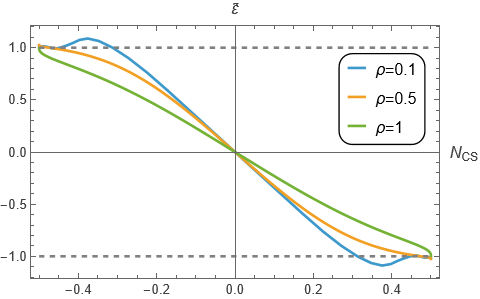}
 		\caption{} \label{eNcs}
 	\end{subfigure}
 	\caption{The normalized fermion energies $\tilde{\varepsilon}$, which was initially assumed to be $x_0$-independent, in terms of (a) Euclidean time $\tilde{x}_0$ and (b) the Chern-Simons number $N_\mathrm{CS}$, for $\tilde{\rho}$=0.1 (blue line), 0.5 (orange line) and 1 (green line).} 
 \end{figure}

We now follow the the usual procedure of solving the system of equations~\eqref{indepeignequ} as an eigenvalue problem for fixed values of $x_0$, and display the results in Fig.~\ref{ex0}. As mentioned earlier, to derive Eq.~\eqref{indepeignequ} we have made the usual assumption that the normalized fermion energy $\tilde{\varepsilon}$ is independent of $\tilde{x}_0$. Contrary to this assumption, Fig.~\ref{ex0} displays the dependence of $\tilde{\varepsilon}$ on $\tilde{x}_0$, whose scale of variation is comparable to that of the profile functions shown in Figs.~(\ref{abcpro})(\ref{hkpro}) and therefore should not be neglected. Moreover, as anticipated from our boundary condition analysis, Fig.~\ref{ex0} confirms that the fermion zero mode occurs for $\tilde{x}_0=0$.  Note that, in this case, the bound states exist only for a finite $x_0$-interval, before joining the continua.
 
 Figure \ref{eNcs} shows the behavior of $\tilde{\varepsilon}$ with respect to $N_\mathrm{CS}$ for $\tilde{\rho}= 0.1,\, 0.5$ and $0.75$, consistent with the observation of Ref.~\cite{Yang:1993qi}.
 As indicated in Fig.~\ref{eNcs}, the value of $\Delta N_\mathrm{CS}$ for level crossing depends on $\tilde{\rho}$: $\Delta N_\mathrm{CS}$ tends to zero as $\tilde{\rho}\rightarrow 0$, while for large $\tilde{\rho}$ it approaches unity.
 
 However, if we consider $\tilde{\varepsilon}$ to be a function of $\tilde{x}_0$ from the beginning, as we should, Eq.~(\ref{indepeignequ}) is modified to
 \begin{subequations}
 	\begin{align}
 		&\tilde{G}_{R}'+(\tilde{h}\tilde{F}_{L}+\tilde{k} \tilde{G}_{L})= (\tilde{x}_0\dot{\tilde{\varepsilon}}+\tilde{\varepsilon})\, \tilde{F}_{R}, \\
 		&\tilde{F}_{R}'+\frac{2}{\tilde{r}} \tilde{F}_{R}-(\tilde{h}\tilde{G}_{L}-\tilde{k}\tilde{F}_{L})= -(\tilde{x}_0\dot{\tilde{\varepsilon}}+\tilde{\varepsilon})\, \tilde{G}_{R}, \\
 		&\tilde{G}_{L}'-2\tilde{a}\tilde{G}_{L}+(2\tilde{b}-\tilde{c})\tilde{F}_{L}-(\tilde{h}\tilde{F}_{R}-\tilde{k}\tilde{G}_{R})=-(\tilde{x}_0\dot{\tilde{\varepsilon}}+\tilde{\varepsilon})\, \tilde{F}_{L}, \\
 		&\tilde{F}_{L}'+(2\tilde{a}+\frac{2}{\tilde{r}})\tilde{F}_{L}+(2\tilde{b}+\tilde{c})\tilde{G}_{L}+(\tilde{k}\tilde{F}_{R}+\tilde{h}\tilde{G}_{R})= (\tilde{x}_0\dot{\tilde{\varepsilon}}+\tilde{\varepsilon})\, \tilde{G}_{L},
 	\end{align} \label{depeignequ}
 \end{subequations}
 which we solve numerically with the mentioned boundary conditions, {\it i.e.,}
 \[\tilde{G}_{L/R}, \tilde{F}_{L/R}\,{\xrightarrow{r\to \infty}}\,0, \qquad \tilde{F}_{L/R}(0)=0.\]

\begin{figure}[ht]
 	\begin{subfigure}{.48\textwidth}
 		\centering
 		\includegraphics[width=1\linewidth]{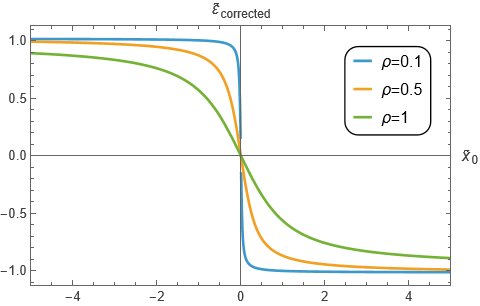}
 		\caption{} \label{etx0}
 	\end{subfigure} \hfill
 	\begin{subfigure}{.49\textwidth}
 		\centering	\includegraphics[width=1\linewidth]{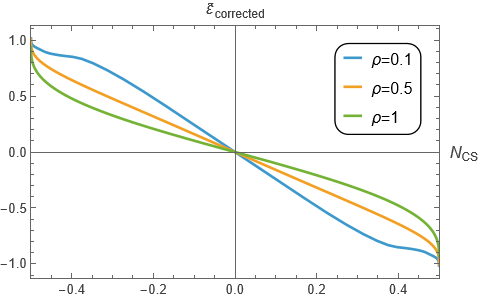}
 		\caption{} \label{etNcs}
 	\end{subfigure}
 	\caption{The normalized fermion energies $\tilde{\varepsilon}$, which is assumed to be $x_0$-independent from the beginning, in terms of (a) Euclidean time $\tilde{x}_0$ and (b) the Chern-Simons number $N_\mathrm{CS}$ for $\tilde{\rho}$=0.1 (blue line), 0.5 (orange line) and 1 (green line).} \label{etx0Ncs}
 \end{figure}

Figure~\ref{etx0Ncs} displays the dependence of corrected energies $\tilde{\varepsilon}$ on (a) $\tilde{x}_0$ and (b) $N_{CS}$ for $\tilde{\rho}= 0.1,\, 0.5$ and $0.75$. Note that the bound states merge with the continua asymptotically at $x_0=\pm \infty$. 
 
\begin{figure}[ht]	
	\centering
	\includegraphics[width=0.49\textwidth]{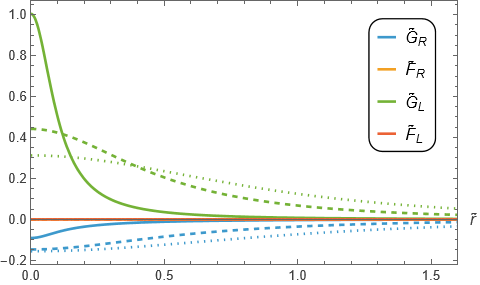}
	\caption{The zero-mode wave functions for $\tilde{\rho}=0.1$ (solid), $\tilde{\rho}=0.5$ (dashed), and $\tilde{\rho}=1$ (dotted).}
	\label{x0infzero}
\end{figure}

Finally, Fig.~\ref{x0infzero} illustrates the fermion zero-mode wave functions for $x_0=0$ with $\tilde{\rho}=0.1$ (solid line), $\tilde{\rho}=0.5$ (dashed line) and $\tilde{\rho}=1$ (dotted line). For this case, the solutions are identical in both the $x_0$-dependent and $x_0$-independent cases. The figure shows that the $F$ functions vanish identically, whereas the $G$ functions are non-vanishing, with the left-handed component being dominant. Moreover, the wave functions become increasingly localized as $\tilde{\rho}$ decreases.

 \section{Conclusion}\label{sec4}
 The fermion level crossing process in the EW instanton background has been studied before, assuming that the fermion energy is Euclidean-time independent. In this approach, the value of $\Delta N_\mathrm{CS}$ seems to depend on $\tilde{\rho}=\rho m_q$, the instanton size times the fermion mass parameter. However, in this work, we have shown that the Euclidean-time dependence of the fermion energy cannot be neglected in comparison with that of the instanton profile functions. Taking into account the Euclidean-time dependence of the energy, we then modified the conventional ansatzes for fermion fields as well as their equations of motion. Finally, by solving these equations numerically, we have reached a qualitatively different result. That is, we have shown that the bound states exit or enter the continua only at $x_0=\pm\infty$, thus verifying the index theorem in the framework of the EW theory: for every, instanton mediated level crossing, $|\Delta N_\mathrm{CS}|=1$, independent of  $\tilde{\rho}$. 
 
 \appendix
 \section{Euclidean formulation of EW} \label{app}
 In this appendix, we describe the procedure for Euclideanizing the EW theory, performing the Wick rotation, 
 \[x^0 = x_0\to-i\hat{x}_0,\qquad x^i=-x_i\to\hat{x}_i,\] 
 where the Euclidean quantities are denoted by a caret and written with lower indices. 
 The corresponding rotation for the gauge field is then defined as \cite{Vainshtein:1981wh}
  \[A_0\to i\hat{A}_0,\qquad A^i=-A_i\to-\hat{A}_i,\] 
 under which the zeroth component of the covariant derivative becomes
 \[D_0=\partial_{0}-ig A_0^a{\tau^a}/{2}\to i\left(\hat{\partial}_{0}-ig\hat{A}_0^a{\tau^a}/{2}\right),\]
 while $D_i$ remains unchanged. Therefore, if we define the following transformation laws for $D_\mu$:
 \begin{align}
 	D_0\to i\hat{D}_0,\qquad D_i=-D^i\to\hat{D}_i,\label{Dhat}
 \end{align}
 the Euclidean covariant derivative can be expressed in terms of $\hat{A}_\mu$ and  $\partial/\partial\hat{x}_{\mu}\,(\hat{\partial}_\mu)$ in the same way as the Minkowskian one. Similarly, for the field strength tensor we need to impose \cite{Vainshtein:1981wh}
 \[F_{0i}^a=-F^{0i,a}\to i\hat{F}_{0i}^a, \quad F_{ij}^a=F^{ij,a}\to \hat{F}_{ij}^a,\]
 so that the invariant combination \[F_{\mu\nu}^a F^{\mu\nu,a}=2F_{0i}^a F^{0i,a}+F_{ij}^a F^{ij,a}\to \hat{F}_{\mu\nu}^a\hat{F}_{\mu\nu}^a,\] as well as \[\mathcal{B}_{\mu\nu}^a \mathcal{B}^{\mu\nu,a}\to \hat{\mathcal{B}}_{\mu\nu}^a\hat{\mathcal{B}}^{\mu\nu,a},\] 
 can be expressed in a form analogous to their Minkowskian counterparts.
 
For the Fermi fields, which we name $\psi$ and $\bar{\psi}$, it is convenient to define \cite{Vainshtein:1981wh}
 \[\psi\to\hat{\psi}, \qquad \bar{\psi}\to-i\hat{\bar{\psi}}.\] 
 Moreover, the four Hermitian $\gamma$-matrices are given in Euclidean space by
 \begin{align*}
 	\hat{\gamma}_0=\begin{pmatrix}
 		0 & 1\\ 1 & 0
 	\end{pmatrix}=\gamma^0,\qquad
 	\hat{\gamma}_i=\begin{pmatrix}
 		0 & -i{\sigma}^i\\ i{\sigma}^i & 0
 	\end{pmatrix}=-i\gamma^i,\qquad
 	\hat{\gamma}_\mu=\begin{pmatrix}
 		0 & \sigma^\mu\\ \bar{\sigma}^\mu & 0
 	\end{pmatrix},
 \end{align*}
 where $\{\sigma^i\}$ denote the Pauli matrices, and $\gamma^5=\gamma^0\gamma^1\gamma^2\gamma^3$. According to these conventions and Eq.~\eqref{Dhat}, we obtain
 \[ i \gamma^{\mu}D_\mu = i\gamma^0 D_0+ i\gamma^i D_i\to\hat{\gamma}_0\hat{D}_0+\hat{\gamma}_i \hat{D}_i=\hat{\gamma}_\mu \hat{D}_\mu,\]
 and thereby, the fermionic terms turn into
 \[\bar{R}^i\, i \gamma^{\mu}\mathcal{D}_\mu R^i\to -\hat{\bar{R}}^i \hat{\gamma}^{\mu}\hat{\mathcal{D}}_\mu \hat{R}^i, \quad \bar{L}^i\, i \gamma^{\mu} D_{\mu}L^i \to -\hat{\bar{L}}^i\hat{\gamma}_\mu \hat{D}_\mu\hat{L}^i.\]
 
Finally, considering $\phi\to-i\hat{\phi}$ for the Higgs field, a typical Yukawa term also transforms as
 \[	-\bar{E}_L^i y_l^{ij}\phi\, e_R^j\to \hat{\bar{E}}_{L}^i y_l^{ij}\hat{\phi}\, \hat{e}_R^j,\]
 and we can write down an expression for the Euclidean action, $\hat{S}=-iS$, 
 in the form of
 \begin{align*}
 	\hat{S}=\int\mathrm{d^4}&\hat{x}  \left[\frac{1}{4}\hat{\mathcal{B}}_{\mu\nu} \hat{\mathcal{B}}_{\mu \nu}+\frac{1}{4}\hat{F}_{\mu \nu}^a \hat{F}_{\mu \nu}^{a}+\hat{D}_{\mu} \hat{\phi}^{\dagger} \hat{D}_{\mu}\hat{\phi}+\lambda\left(\hat{\phi}^{\dagger} \hat{\phi} -\frac{v^{2}}{2} \right)^{2} \right. \\ & \left.  +\hat{\bar{R}}^i \hat{\gamma}^{\mu}\hat{\mathcal{D}}_\mu \hat{R}^i+\hat{\bar{L}}^i \hat{\gamma}^{\mu} \hat{D}_{\mu}\hat{L}^i  
 	-\hat{\bar{E}}_{L}^i y_l^{ij}\hat{\phi}\, \hat{e}_R^j - \hat{\bar{Q}}_{L}^iy_q^{ij}\left(\hat{\tilde{\phi}}\hat{u}_{R}^j+\hat{\phi} \hat{d}_{R}^j\right)+\mathrm{H.C.} \right].
 \end{align*}
 Throughout this paper, we have used the Euclidean quantities, omitting the carets. 
 
\bibliographystyle{ieeetr}
\bibliography{bib}

@article{JackiwSchrieffer1981,
	author = {Jackiw, R. and Schrieffer, J. R.},
	title = {Solitons with Fermion Number $1/2$ in Condensed Matter and Relativistic Field Theories},
	journal = {Nuclear Physics B},
	volume = {190},
	number = {2},
	pages = {253--265},
	year = {1981},
	doi = {10.1016/0550-3213(81)90044-4}
}

@article{SuSchriefferHeeger1979,
	author = {Su, W. P. and Schrieffer, J. R. and Heeger, A. J.},
	title = {Solitons in Polyacetylene},
	journal = {Physical Review Letters},
	volume = {42},
	number = {25},
	pages = {1698--1701},
	year = {1979},
	doi = {10.1103/PhysRevLett.42.1698}
}

@article{Kitaev2001,
	author = {Kitaev, A. Yu.},
	title = {Unpaired Majorana Fermions in Quantum Wires},
	journal = {Physics-Uspekhi},
	volume = {44},
	pages = {131--136},
	year = {2001},
	doi = {10.1070/1063-7869/44/10S/S29}
}

@book{rajaraman1982solitons,
	title={Solitons and Instantons: An Introduction to Solitons and Instantons in Quantum Field Theory},
	author={Rajaraman, R.},
	isbn={9780444862297},
	lccn={lc81022510},
	series={North-Holland personal library},
	url={https://books.google.com/books?id=1XucQgAACAAJ},
	year={1982},
	publisher={North-Holland Publishing Company}
}

@article{Yang:1993qi,
	author = "Yang, Ke-yan",
	title = "{The Process of fermion level crossing in electroweak instanton}",
	reportNumber = "DPNU-93-42",
	doi = "10.1103/PhysRevD.49.5491",
	journal = "Phys. Rev. D",
	volume = "49",
	pages = "5491--5496",
	year = "1994"
}

@book{manton2004topological,
	title={Topological Solitons},
	author={Manton, N. and Sutcliffe, P.},
	isbn={9780521838368},
	lccn={2003069072},
	series={Cambridge Monographs on Mathematical Physics},
	url={https://books.google.com/books?id=DptFmAEACAAJ},
	year={2004},
	publisher={Cambridge University Press}
}

@article{tHooft:1976rip,
	author = "'t Hooft, Gerard",
	editor = "Shifman, Mikhail A.",
	title = "{Symmetry Breaking Through Bell-Jackiw Anomalies}",
	reportNumber = "PRINT-76-0254 (HARVARD)",
	doi = "10.1103/PhysRevLett.37.8",
	journal = "Phys. Rev. Lett.",
	volume = "37",
	pages = "8--11",
	year = "1976"
}

@article{tHooft:1976snw,
	author = "'t Hooft, Gerard",
	editor = "Shifman, Mikhail A.",
	title = "{Computation of the Quantum Effects Due to a Four-Dimensional Pseudoparticle}",
	reportNumber = "PRINT-76-0551 (HARVARD)",
	doi = "10.1103/PhysRevD.14.3432",
	journal = "Phys. Rev. D",
	volume = "14",
	pages = "3432--3450",
	year = "1976",
	note = "[Erratum: Phys.Rev.D 18, 2199 (1978)]"
}

@article{Affleck:1980mp,
	author = "Affleck, Ian",
	editor = "Shifman, Mikhail A.",
	title = "{On Constrained Instantons}",
	reportNumber = "HUTP-80-A075",
	doi = "10.1016/0550-3213(81)90307-2",
	journal = "Nucl. Phys. B",
	volume = "191",
	pages = "429",
	year = "1981"
}

@article{Derrick:1964ww, author = "Derrick, G. H.", title = "{Comments on nonlinear wave equations as models for elementary particles}", doi = "10.1063/1.1704233", journal = "J. Math. Phys.", volume = "5", pages = "1252--1254", year = "1964" }

@article{Gibbs:1995xt, author = "Gibbs, M. J.", title = "{Electroweak instantons at nonzero Weinberg angle}", eprint = "hep-ph/9501253", archivePrefix = "arXiv", reportNumber = "CAVENDISH-HEP-94-5", doi = "10.1016/0370-2693(95)00126-6", journal = "Phys. Lett. B", volume = "348", pages = "149--154", year = "1995" 
}

@article{d370fac2b94a4edca23e7cf28a022b27,
title = "Pseudoparticle solutions of the Yang-Mills equations",
author = "Belavin, {A. A.} and Polyakov, {A. M.} and Schwartz, {A. S.} and Tyupkin, {Yu S.}",
year = "1975",
month = oct,
day = "13",
doi = "10.1016/0370-2693(75)90163-X",
language = "English (US)",
volume = "59",
pages = "85--87",
journal = "Physics Letters B",
issn = "0370-2693",
publisher = "Elsevier",
number = "1",
}

@article{Brown:1977eb,
    author = "Brown, Lowell S. and Carlitz, Robert D. and Creamer, Dennis B. and Lee, Choon-kyu",
    editor = "Shifman, Mikhail A.",
    title = "{Propagation Functions in Pseudoparticle Fields}",
    reportNumber = "RLO-1388-735",
    doi = "10.1103/PhysRevD.17.1583",
    journal = "Phys. Rev. D",
    volume = "17",
    pages = "1583",
    year = "1978"
}

@article{Shahkarami_2011,
   title={Exact solutions of a fermion-soliton system in two dimensions},
   volume={2011},
   ISSN={1029-8479},
   url={http://dx.doi.org/10.1007/JHEP06(2011)116},
   DOI={10.1007/jhep06(2011)116},
   number={6},
   journal={Journal of High Energy Physics},
   publisher={Springer Science and Business Media LLC},
   author={Shahkarami, L. and Gousheh, S. S.},
   year={2011},
   month=jun }

@article{Kunz_1994,
	title={Level crossing along sphaleron barriers},
	volume={50},
	ISSN={0556-2821},
	url={http://dx.doi.org/10.1103/PhysRevD.50.1051},
	DOI={10.1103/physrevd.50.1051},
	number={2},
	journal={Physical Review D},
	publisher={American Physical Society (APS)},
	author={Kunz, Jutta and Brihaye, Yves},
	year={1994},
	month=jul, pages={1051–1059} }

@article{Jackiw:1976pf,
	author         = {Jackiw, R. and Rebbi, C.},
	title          = {Vacuum Periodicity in a Yang-Mills Quantum Theory},
	journal        = {Phys. Rev. Lett.},
	volume         = {37},
	pages          = {172--175},
	year           = {1976},
	doi            = {10.1103/PhysRevLett.37.172}
}

@article{Callan:1976je,
	author         = {Callan, Curtis G., Jr. and Dashen, Roger F. and Gross, David J.},
	title          = {The Structure of the Gauge Theory Vacuum},
	journal        = {Phys. Lett. B},
	volume         = {63},
	pages          = {334--340},
	year           = {1976},
	doi            = {10.1016/0370-2693(76)90277-X}
}

@article{Klinkhamer:1984di,
	author         = {Klinkhamer, F. R. and Manton, N. S.},
	title          = {A Saddle-Point Solution in the Weinberg-Salam Theory},
	journal        = {Phys. Rev. D},
	volume         = {30},
	pages          = {2212},
	year           = {1984},
	doi            = {10.1103/PhysRevD.30.2212}
}

@article{PhysRevD.45.2920,
	title = {Skyrmions, vacuum polarization, and the adiabatic method},
	author = {Gousheh, Siamak S.},
	journal = {Phys. Rev. D},
	volume = {45},
	issue = {8},
	pages = {2920--2928},
	numpages = {0},
	year = {1992},
	month = {Apr},
	publisher = {American Physical Society},
	doi = {10.1103/PhysRevD.45.2920},
	url = {https://link.aps.org/doi/10.1103/PhysRevD.45.2920}
}

@article{Atiyah:1963zz,
	author = "Atiyah, M. F. and Singer, I. M.",
	title = "{The index of elliptic operators on compact manifolds}",
	doi = "10.1090/S0002-9904-1963-10957-X",
	journal = "Bull. Am. Math. Soc.",
	volume = "69",
	pages = "422--433",
	year = "1969"
}

@article{JackiwRebbi1976,
	author       = {R. Jackiw and C. Rebbi},
	title        = {Solitons with Fermion Number 1/2},
	journal      = {Phys. Rev. D},
	volume       = {13},
	pages        = {3398--3409},
	year         = {1976},
	doi          = {10.1103/PhysRevD.13.3398},
	note         = {Seminal work demonstrating fermion zero modes in soliton/monopole backgrounds.}
}

@article{Polyakov:1974ek,
	author = "Polyakov, Alexander M.",
	editor = "Taylor, J. C.",
	title = "{Particle Spectrum in Quantum Field Theory}",
	reportNumber = "PRINT-74-1566 (LANDAU-INST)",
	journal = "JETP Lett.",
	volume = "20",
	pages = "194--195",
	year = "1974"
}

@article{tHooft:1974kcl,
	author = "'t Hooft, Gerard",
	editor = "Taylor, J. C.",
	title = "{Magnetic Monopoles in Unified Gauge Theories}",
	reportNumber = "CERN-TH-1876",
	doi = "10.1016/0550-3213(74)90486-6",
	journal = "Nucl. Phys. B",
	volume = "79",
	pages = "276--284",
	year = "1974"
}

@article{Adler1969,
	author = {Adler, Stephen L.},
	title = {Axial-Vector Vertex in Spinor Electrodynamics},
	journal = {Physical Review},
	volume = {177},
	number = {5},
	pages = {2426--2438},
	year = {1969},
	doi = {10.1103/PhysRev.177.2426}
}

@article{BellJackiw1969,
	author = {Bell, J. S. and Jackiw, R.},
	title = {A PCAC Puzzle: $\pi^0 \to \gamma\gamma$ in the $\sigma$ Model},
	journal = {Il Nuovo Cimento A},
	volume = {60},
	number = {1},
	pages = {47--61},
	year = {1969},
	doi = {10.1007/BF02823296}
}

@article{Manton1983,
	author = {Manton, N. S.},
	title = {Topology in the Weinberg--Salam Theory},
	journal = {Physical Review D},
	volume = {28},
	number = {8},
	pages = {2019--2026},
	year = {1983},
	doi = {10.1103/PhysRevD.28.2019}
}

@article{GoldstoneWilczek1981,
	author = {Goldstone, Jeffrey and Wilczek, Frank},
	title = {Fractional Quantum Numbers on Solitons},
	journal = {Physical Review Letters},
	volume = {47},
	number = {14},
	pages = {986--989},
	year = {1981},
	doi = {10.1103/PhysRevLett.47.986}
}

@article{PhysRevD.26.2058,
	title = {Dyon-fermion dynamics},
	author = {Callan, Curtis G.},
	journal = {Phys. Rev. D},
	volume = {26},
	issue = {8},
	pages = {2058--2068},
	numpages = {0},
	year = {1982},
	month = {Oct},
	publisher = {American Physical Society},
	doi = {10.1103/PhysRevD.26.2058},
	url = {https://link.aps.org/doi/10.1103/PhysRevD.26.2058}
}

@article{Kuzmin:1985mm,
	author = "Kuzmin, V. A. and Rubakov, V. A. and Shaposhnikov, M. E.",
	title = "{On the Anomalous Electroweak Baryon Number Nonconservation in the Early Universe}",
	reportNumber = "IC/85/8",
	doi = "10.1016/0370-2693(85)91028-7",
	journal = "Phys. Lett. B",
	volume = "155",
	pages = "36",
	year = "1985"
}

@article{Ringwald:1989ee,
	author = "Ringwald, A.",
	title = "{High-Energy Breakdown of Perturbation Theory in the Electroweak Instanton Sector}",
	reportNumber = "DESY-89-074",
	doi = "10.1016/0550-3213(90)90300-3",
	journal = "Nucl. Phys. B",
	volume = "330",
	pages = "1--18",
	year = "1990"
}

@article{PhysRevD.36.581,
	title = {Sphalerons, small fluctuations, and baryon-number violation in electroweak theory},
	author = {Arnold, Peter and McLerran, Larry},
	journal = {Phys. Rev. D},
	volume = {36},
	issue = {2},
	pages = {581--595},
	numpages = {0},
	year = {1987},
	month = {Jul},
	publisher = {American Physical Society},
	doi = {10.1103/PhysRevD.36.581},
	url = {https://link.aps.org/doi/10.1103/PhysRevD.36.581}
}

@article{Vainshtein:1981wh,
	author = "Vainshtein, A. I. and Zakharov, Valentin I. and Novikov, V. A. and Shifman, Mikhail A.",
	title = "{ABC's of Instantons}",
	reportNumber = "ITEP-2-1981",
	doi = "10.1070/PU1982v025n04ABEH004533",
	journal = "Sov. Phys. Usp.",
	volume = "25",
	pages = "195",
	year = "1982"
}
\end{document}